\documentclass[11pt,fleqn]{book} 

\usepackage{tcolorbox}
\usepackage{wrapfig}
\usepackage{url}
\usepackage[utf8]{inputenc}
\usepackage{aas_macros}
\usepackage{pdfpages}
\usepackage{multicol}
\usepackage{setspace}
\usepackage{tcolorbox}


\usepackage[top=2.54cm,bottom=2.54cm,left=2.54cm,right=2.54cm,headsep=10pt,letterpaper]{geometry} 
\usepackage{xcolor} 
\definecolor{ocre}{RGB}{52,177,201} 

\usepackage{avant} 
\usepackage{mathptmx} 

\usepackage{microtype} 
\usepackage[utf8]{inputenc} 
\usepackage[T1]{fontenc} 

\usepackage{titlesec} 

\usepackage{graphicx} 
\graphicspath{{Pictures/}{figures/cosmo/}} 

\usepackage{lipsum} 

\usepackage{tikz} 

\usepackage[english]{babel} 

\usepackage{enumitem} 
\setlist{nolistsep} 

\usepackage{booktabs} 

\usepackage{eso-pic} 

\usepackage{titletoc} 

\contentsmargin{0cm} 
\titlecontents{chapter}[1.25cm] 
{\addvspace{15pt}\Large\sffamily\bfseries} 
{\color{ocre!60}\contentslabel[\LARGE\thecontentslabel]{1.25cm}\color{ocre}} 
{}
{\color{ocre!60}\normalsize\sffamily\bfseries\;\titlerule*[.5pc]{.}\;\thecontentspage} 
\titlecontents{section}[1.25cm] 
{\addvspace{5pt}\sffamily\bfseries} 
{\contentslabel[\thecontentslabel]{1.25cm}} 
{}
{\sffamily\hfill\color{black}\thecontentspage} 
[]
\titlecontents{subsection}[1.25cm] 
{\addvspace{1pt}\sffamily\small} 
{\contentslabel[\thecontentslabel]{1.25cm}} 
{}
{\sffamily\;\titlerule*[.5pc]{.}\;\thecontentspage} 
[]

\titlecontents{lsection}[0em] 
{\footnotesize\sffamily} 
{}
{}
{}

\titlecontents{lsubsection}[.5em] 
{\normalfont\footnotesize\sffamily} 
{}
{}
{}

\usepackage{fancyhdr} 

\fancypagestyle{plain}{\fancyhead{}} 

\makeatletter
\renewcommand{\cleardoublepage}{
\clearpage\ifodd\c@page\else
\hbox{}
\vspace*{\fill}
\thispagestyle{empty}
\fi}

\usepackage{amsmath,amsfonts,amssymb,amsthm} 

\newtheoremstyle{ocrenumbox}
{0pt}
{0pt}
{\normalfont}
{}
{\small\bf\sffamily\color{ocre}}
{\;}
{0.25em}
{\small\sffamily\color{ocre}\thmname{#1}\nobreakspace\thmnumber{\@ifnotempty{#1}{}\@upn{#2}}
\thmnote{\nobreakspace\the\thm@notefont\sffamily\bfseries\color{black}---\nobreakspace#3.}} 

\newtheoremstyle{blacknumex}
{5pt}
{5pt}
{\normalfont}
{} 
{\small\bf\sffamily}
{\;}
{0.25em}
{\small\sffamily{\tiny\ensuremath{\blacksquare}}\nobreakspace\thmname{#1}\nobreakspace\thmnumber{\@ifnotempty{#1}{}\@upn{#2}}
\thmnote{\nobreakspace\the\thm@notefont\sffamily\bfseries---\nobreakspace#3.}}

\newtheoremstyle{blacknumbox} 
{0pt}
{0pt}
{\normalfont}
{}
{\small\bf\sffamily}
{\;}
{0.25em}
{\small\sffamily\thmname{#1}\nobreakspace\thmnumber{\@ifnotempty{#1}{}\@upn{#2}}
\thmnote{\nobreakspace\the\thm@notefont\sffamily\bfseries---\nobreakspace#3.}}

\newtheoremstyle{ocrenum}
{5pt}
{5pt}
{\normalfont}
{}
{\small\bf\sffamily\color{ocre}}
{\;}
{0.25em}
{\small\sffamily\color{ocre}\thmname{#1}\nobreakspace\thmnumber{\@ifnotempty{#1}{}\@upn{#2}}
\thmnote{\nobreakspace\the\thm@notefont\sffamily\bfseries\color{black}---\nobreakspace#3.}} 
\makeatother

\newcounter{dummy}
\numberwithin{dummy}{section}
\theoremstyle{ocrenumbox}
\newtheorem{theoremeT}[dummy]{Theorem}

\newtheorem{exerciseT}{Exercise}[chapter]
\theoremstyle{blacknumex}
\newtheorem{exampleT}{Example}[chapter]
\theoremstyle{blacknumbox}

\newtheorem{definitionT}{Definition}[section]
\newtheorem{corollaryT}[dummy]{Corollary}
\theoremstyle{ocrenum}

\RequirePackage[framemethod=default]{mdframed} 

\newmdenv[skipabove=7pt,
skipbelow=7pt,
backgroundcolor=black!5,
linecolor=ocre,
innerleftmargin=5pt,
innerrightmargin=5pt,
innertopmargin=5pt,
leftmargin=0cm,
rightmargin=0cm,
innerbottommargin=5pt]{tBox}

\newmdenv[skipabove=7pt,
skipbelow=7pt,
rightline=false,
leftline=true,
topline=false,
bottomline=false,
backgroundcolor=ocre!10,
linecolor=ocre,
innerleftmargin=5pt,
innerrightmargin=5pt,
innertopmargin=5pt,
innerbottommargin=5pt,
leftmargin=0cm,
rightmargin=0cm,
linewidth=4pt]{eBox}	

\newmdenv[skipabove=7pt,
skipbelow=7pt,
rightline=false,
leftline=true,
topline=false,
bottomline=false,
linecolor=ocre,
innerleftmargin=5pt,
innerrightmargin=5pt,
innertopmargin=0pt,
leftmargin=0cm,
rightmargin=0cm,
linewidth=4pt,
innerbottommargin=0pt]{dBox}	

\newmdenv[skipabove=7pt,
skipbelow=7pt,
rightline=false,
leftline=true,
topline=false,
bottomline=false,
linecolor=gray,
backgroundcolor=black!5,
innerleftmargin=5pt,
innerrightmargin=5pt,
innertopmargin=5pt,
leftmargin=0cm,
rightmargin=0cm,
linewidth=4pt,
innerbottommargin=5pt]{cBox}

\makeatletter
\renewcommand{\@seccntformat}[1]{\llap{\textcolor{ocre}{\csname the#1\endcsname}\hspace{1em}}}
\renewcommand{\section}{\@startsection{section}{1}{\z@}
{-2ex \@plus -1ex \@minus -.2ex}
{1ex \@plus.1ex }
{\normalfont\large\sffamily\bfseries}}
\renewcommand{\subsection}{\@startsection {subsection}{2}{\z@}
{-2ex \@plus -0.1ex \@minus -.2ex}
{0.5ex \@plus.2ex }
{\normalfont\sffamily\bfseries}}
\renewcommand{\subsubsection}{\@startsection {subsubsection}{3}{\z@}
{-2ex \@plus -0.1ex \@minus -.2ex}
{.2ex \@plus.2ex }
{\normalfont\small\sffamily\bfseries}}
\renewcommand\paragraph{\@startsection{paragraph}{4}{\z@}
{-2ex \@plus-.2ex \@minus .2ex}
{.1ex}
{\normalfont\small\sffamily\bfseries}}

\usepackage[hidelinks,backref=page,pagebackref=true,hyperindex=true,colorlinks=false,breaklinks=true,urlcolor= ocre,bookmarks=true,bookmarksopen=false,pdftitle={Title},pdfauthor={Author}]{hyperref}

\newcommand{\thechapterimage}{}
\newcommand{\chapterimage}[1]{\renewcommand{\thechapterimage}{#1}}

\def\thechapter{\arabic{chapter}}
\def\@makechapterhead#1{
\thispagestyle{empty}
{\centering \normalfont\sffamily
\ifnum \c@secnumdepth >\m@ne
\if@mainmatter
\startcontents
\begin{tikzpicture}[remember picture,overlay]
\node at (current page.north west)
{\begin{tikzpicture}[remember picture,overlay]
\node[anchor=north west,inner sep=0pt] at (0,0) {\includegraphics[width=\paperwidth]{\thechapterimage}};
\draw[anchor=west] (2.46cm,-6cm) node [rounded corners=20pt,fill=ocre!10!white,text opacity=10,draw=ocre,draw opacity=1,line width=1.5pt,fill opacity=.6,inner sep=12pt]{\LARGE\sffamily\bfseries\textcolor{black}{\thechapter. #1\strut\makebox[22cm]{}}};
\end{tikzpicture}};
\end{tikzpicture}}
\par\vspace*{130\p@}
\fi
\fi}

\def\@makeschapterhead#1{
\thispagestyle{empty}
{\centering \normalfont\sffamily
\ifnum \c@secnumdepth >\m@ne
\if@mainmatter
\begin{tikzpicture}[remember picture,overlay]
\node at (current page.north west)
{\begin{tikzpicture}[remember picture,overlay]
\node[anchor=north west,inner sep=0pt] at (0,0) {\includegraphics[width=\paperwidth]{\thechapterimage}};
\draw[anchor=west] (2.46cm,-6cm) node [rounded corners=20pt,fill=ocre!10!white,fill opacity=.6,inner sep=12pt,text opacity=1,draw=ocre,draw opacity=1,line width=1.5pt]{\LARGE\sffamily\bfseries\textcolor{black}{#1\strut\makebox[22cm]{}}};
\end{tikzpicture}};
\end{tikzpicture}}
\par\vspace*{120\p@}
\fi
\fi
}
\makeatother

\usepackage{graphicx}
\graphicspath{{./figures/}}
\usepackage{appendix}
\usepackage{latexsym,amsmath,amsfonts,amssymb,booktabs}
\usepackage[font=small]{caption}
\usepackage{slashed,upgreek,amscd,cancel,tensor,color}
\usepackage{adjustbox}
\usepackage[numbers,sort&compress,square]{natbib}
\usepackage{epsfig,latexsym}
\usepackage{url}
\numberwithin{equation}{section}
\usepackage{doi}
\usepackage{subcaption}
\usepackage{mathtools}
\usepackage{upgreek}
\usepackage{stfloats}
\usepackage{afterpage}
\usepackage{multirow}

\begin{document}
\pagenumbering{roman}

\includepdf{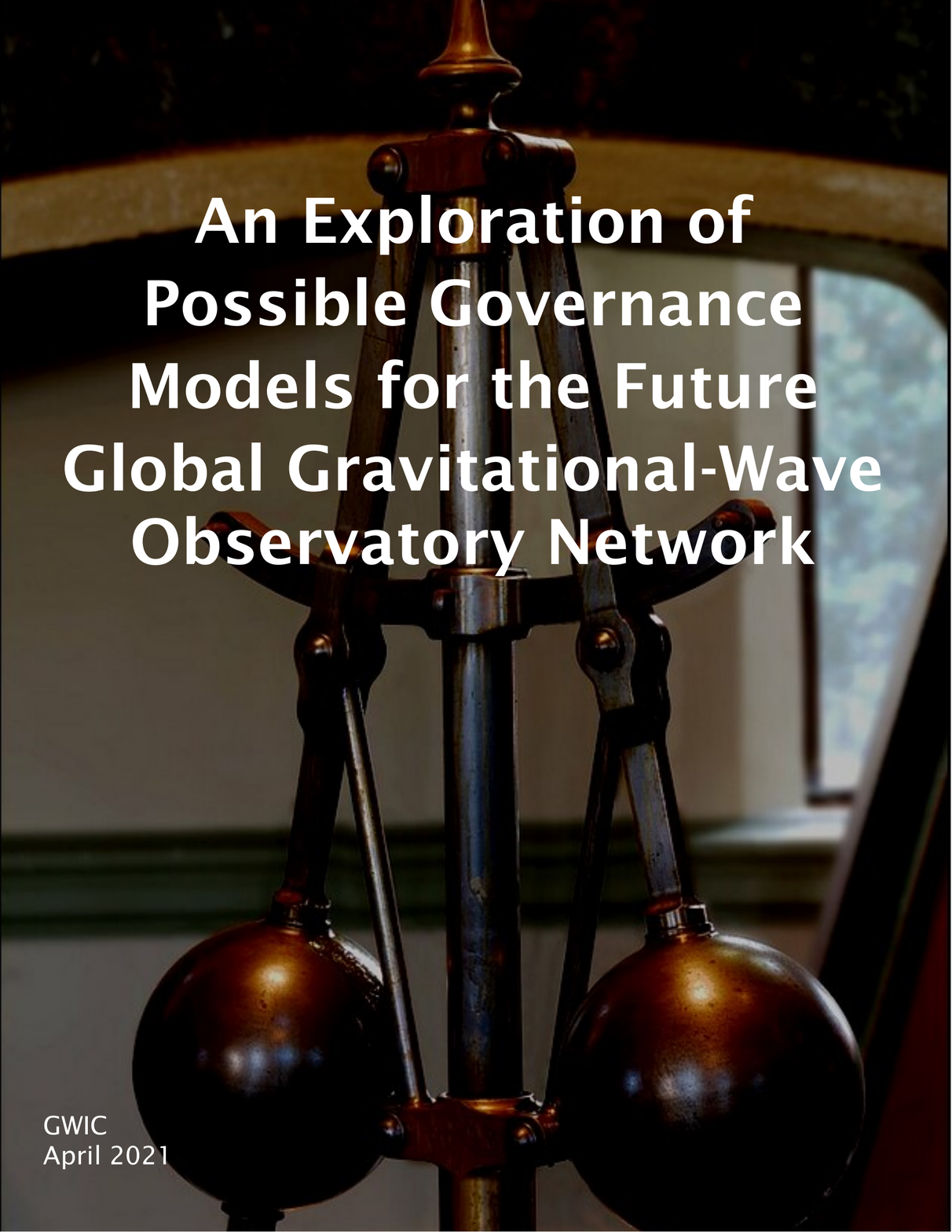}

\begingroup
\thispagestyle{empty}



\newpage
\thispagestyle{empty}

\textbf{GOVERNANCE MODELS SUBCOMMITTEE}

Stavros Katsanevas, European Gravitational Observatory, Italy (Co-chair)

Gary Sanders, Caltech, USA (Co-chair)

Beverly Berger, Stanford University, USA

Gabriela González, Louisiana State University, USA

James Hough, University of Glasgow, UK

Ajit K. Kembhavi, Inter-University Centre for Astronomy and Astrophysics, India

David McClelland, Australian National University, Australia

Masatake Ohashi, Institute of Cosmic Ray Research, Japan

Fulvio Ricci, Universit\`a La Sapienza, and INFN Roma, Italy

Stan Whitcomb, Caltech, USA\\

\textbf{STEERING COMMITTEE}

Michele Punturo, INFN Perugia, Italy (Co-chair)

David Reitze, Caltech, USA (Co-chair)

Peter Couvares, Caltech, USA

Stavros Katsanevas, European Gravitational Observatory

Takaaki Kajita, University of Tokyo, Japan

Vicky Kalogera, Northwestern University, USA

Harald Lueck, AEI, Hannover, Germany

David McClelland, Australian National University, Australia

Sheila Rowan, University of Glasgow, UK

Gary Sanders, Caltech, USA

B.S. Sathyaprakash, Penn State University, USA and Cardiff University, UK

David Shoemaker, MIT, USA (Secretary)

Jo van den Brand, Nikhef, Netherlands
\vspace{1.5cm}


\noindent \textsc{Gravitational Wave International Committee}\\

\noindent{This document was produced by the GWIC 3G Subcommittee and the GWIC 3G Governance Models Subcommittee}\\ 

\noindent \textit{Final release, April 2021}\\ 

\noindent \textit{Cover: Thomas Coxon. Flikr and Creative Commons}


\chapterimage{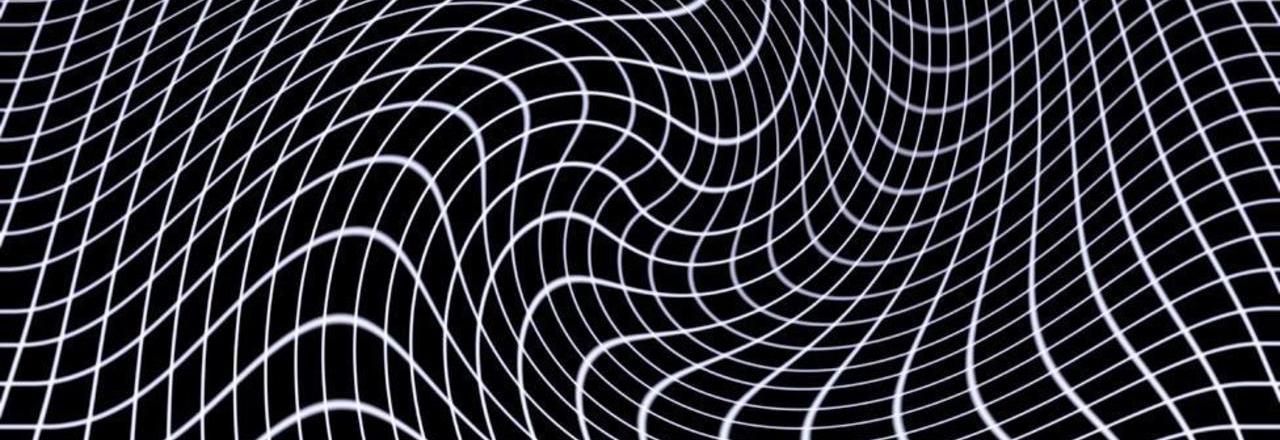} 
\pagestyle{empty} 
\tableofcontents 


\pagestyle{fancy} 
\newpage


\pagenumbering{arabic}
\chapterimage{Ultimate_Goals_pic.jpg} 
\chapter{Ultimate Goals}
\label{ch:introduction}

\vspace{0.5cm}
{\Large\bf I}n the next few years advanced gravitational‐wave (GW) observatories, such as LIGO in the USA, Virgo in Europe, KAGRA in Japan and LIGO‐India in India, will detect hundreds of gravitational‐wave signals emitted by various sources such as those already seen due to the coalescence of binary systems. These instruments, usually referred to as second generation, or 2G detectors, are opening the era of gravitational wave astronomy, just as Galileo’s optical telescope opened the field of optical astronomy more than 400 years ago.
Research and development is underway to design third‐generation (3G) instruments with the sensitivity to observe coalescences of black‐hole black‐hole (BH‐BH) binary systems up to distances corresponding to redshifts 10 and above.

Already, the first observations have demonstrated that full exploration of the GW sky requires a well‐ coordinated network of GW detectors located on different continents: this is a crucial requirement for source localization and polarization studies needed for precision gravitational wave astronomy.

Source location accuracy of gravitational wave data is a limiting factor for many scientific goals of gravitational‐wave astronomy, including especially coordination of multi‐messenger observations. Multi‐ continent networks inherently bring long baselines and three‐dimensional orientation coverage to enable more precise source locations and measurement of gravitational wave polarization. This enables clean reconstruction of source coordinates and GW waveforms, providing unique information concerning the astrophysical processes underlying the sources. This aspect is even more crucial for phenomena, as for example, supernovae explosions for which the waveform prediction has a high intrinsic uncertainty due to the complexity of the physical processes involved. Further, coordination with space GW observatories (e.g. LISA) can bring more precise measurements of the inspiral phase and/or sufficient early warning.1

The combination of measurements from several detectors increases the detection confidence for faint signals emitted even in the deep cosmos, minimizing any instrumental and /or environmental artifacts. A global network also provides for robustness against local disturbances (e.g., storms, power outages) that can disrupt sensitive observations at individual locations.

The construction of a global network of detectors is the cornerstone to scientific success for 3rd generation gravitational wave detectors. Moreover, if carried out with a vision to the future, the third generation implementation, in its infrastructure, technology base and governance can provide the point of departure for subsequent developments beyond the third generation, a path to the long and revolutionary future of gravitational wave astrophysics. The goal of this document is to lay out considerations that influence the optimal choice of governance and to lay out a possible path that can lead the community to an optimal governance model.

\chapterimage{Starting_Points_pic.jpg} 
\chapter{Starting Points and Initial Conditions}
\label{ch:ns}

{\Large\bf T}he current governance structures of the various gravitational wave projects, and particularly the agreements among them to collaborate, have grown up through an ad hoc process and without systematic consideration of the needs for network operation.\\

\begin{itemize}
\item LIGO was initially constructed and operated from a university‐based laboratory via a cooperative agreement from the US National Science Foundation. Advanced LIGO included significant additional contributions from the UK, Germany and Australia, through MOUs with the LIGO Laboratory.
\vspace{1mm}
\item Virgo was constructed by a collaboration of Italian and French groups and operated through the EGO consortium with founding members INFN and CNRS. Netherlands later joined the consortium as an observer, providing significant in‐kind contributions to the construction of Advanced Virgo. Groups from Poland, Hungary and Spain later joined the Virgo collaboration.
\vspace{1mm}
\item KAGRA has recently completed construction and operates under an agreement with its funding agency MEXT. International partners have been recruited, and these make in‐kind contribution through MOU‐style agreements.
\vspace{1mm}
\item The LIGO‐India Project is working towards construction of LIGO‐India with selection of its site, under an agreement with the Indian Department of Atomic Energy (DAE) and the Department of Science and Technology (DST) with the cooperative support of LIGO and the US National Science Foundation.\\
\end{itemize}
\vspace{-2mm}

In all these cases, the governance structures and the relationship to the majority funding agency (or in the case of Virgo, the two largest funding agencies) are the normal ones for a project operating entirely within that country/region. Decisions about the design of the different detectors, scheduling of construction and operation are made by individual projects with approval from the majority funding agencies. The participation by the minority parties grants scientific privileges and rights, but limited roles in the project governance.

In 2006, LIGO and Virgo created the beginnings of an international network, as a collaboration of collaborations. Their agreement, a project‐to‐project Memorandum of Agreement (MoA), created an integrated data analysis effort, with unified management structures for the data analysis groups. Under this MoA, all data taken by both collaborations are analyzed as a single dataset, and all full collaboration results are published jointly, regardless of which detector(s) the data came from.  They are also presently in the process of unifying their computing framework, inside a scheme called IGWN (International Gravitational Wave Observatory Network). The MoA was recently extended to include KAGRA. The first detections of gravitational waves from BH‐BH and NS‐NS inspiral/merger events have been a great success for these earlier collaborations:\\
    \begin{itemize}
        \item The first detection of gravitational waves, made by the LIGO and Virgo collaborations using LIGO data, benefited greatly from in‐depth scrutiny by experts from both LIGO and Virgo. In the view of many of the participants, this critical internal examination was one of the main factors behind the rapid acceptance of this remarkable claim.
        \item The first detection of a binary neutron star (BNS) merger and its localization on the sky via triangulation used data from the two LIGO and single Virgo detectors.\\
    \end{itemize}

In general, the joint LIGO/Virgo management structures for the data analysis effort served the collaborations and the community adequately through these early successes, but the resilience of the joint agreement was not tested by the level of challenge expected in the construction and commissioning of a major new facility.

The LIGO‐Virgo MoA also encourages collaboration on broader matters such as detector R\&D, upgrades and run scheduling, but the details are rather nonspecific. Ad hoc joint experimental working groups have facilitated communication of plans and results among the different projects, but joint activities in the experimental arena have been only a small part of the overall effort. The MoA creates a committee to coordinate observing runs, but this committee has limited power to control the plans of the separate projects. The commitment to joint runs is only voluntary, and the full power of the collaborations has not been brought to bear to maximize joint observing. Despite the informal and collegial nature of the arrangement, it has been remarkably successful due to the huge effort to implement and operate the detectors involving scores of scientists and engineers. What is arguably the most compelling justification for a 3G network, the localization of the BNS merger, was a lucky chance, occurring in the one month of joint running that LIGO and Virgo managed to schedule in August 2017. While historically successful, it should not be used as an argument that the style of collaboration between LIGO and Virgo is sufficient as a model for the future.

Looking to the future, the experimental community, led in large measure by experimenters from the LIGO and Virgo scientific collaborations, has been developing ideas for even more capable detectors. These include improvements/upgrades to current LIGO and Virgo detectors evolving over a decade or more. However, the largest gains in sensitivity and frequency coverage will come from 3rd generation detectors in new, more capable facilities. Several concepts have been developed with varying degrees of depth. However, most of these concepts have been developed with a regional perspective in terms of organization and technical implementation, rather than a global one. These early steps make it appear that the community tends to follow the earlier path of building regional collaborations that are then linked together in the collaboration of collaborations model. Moreover, any rapidly developing regional initiative should be considered in the light of developing the optimal, future international network.

It is important for the gravitational wave community to consider what the right course is, given the current starting point. The larger facilities and more sophisticated detectors of the 3G network are likely to cost \$1B or substantially more per site, a considerable increase over the current facilities/detectors. Funding of any billion dollar, or multi‐billion dollar, scale will require equally significant discussion and support within the larger astronomy and astrophysics communities. At the same time, different decision paths followed in the various nations and regions will compete with the broader global discussion.

Within this tension, there will have to be some respect for the definition of the regional roadmaps, in which the regional priorities for future large research infrastructures are defined. These have to overlap and merge with the global community priorities.4 The growing importance of multi‐messenger astronomy for the 3G network, as well as the coordination with LISA and other space projects will place an even greater premium on coincident running, and the governance structures for it must be sufficient to assure that this will be optimized within the constraints of funding and practicality.

Finally, the governance structures for a 3G gravitational wave network must be acceptable to the relevant funding agencies, and this provides another set of boundary conditions. In particular, an international network will by necessity have to maintain relations with a number of national funding agencies, and this will only be practical if the funding agencies also have an active and engaged parallel forum for discussing plans and for making common decisions. Two such organizations have come into existence in the past decade, with possible relevance to 3G gravitational wave networks:\\

    \begin{itemize}
        \item Committee (ApPIC) of the International Union of Pure and Applied Physics (IUPAP). ApPEC has a similar mandate as APIF but with a more limited geographic scope. As a Working Group of IUPAP, ApPIC is composed of active scientists and leaders in the broader field of astroparticle field, but it does not include ground‐based gravitational wave detectors within its scope, since this is covered primarily in GWIC (a parallel Working Group in the IUPAP hierarchy). APIF was renewed under the aegis of OECD‐GSF for a second three‐year term in 2013, with the expectation that it would become independent after 2016. APIF then drafted the “Astroparticle Physics International Forum Agreement”, which was adopted by its members, and it continues to provide a discussion forum for its members, as an independent body.
\vspace{1mm}
        \item The Gravitational Wave Agencies Correspondents (GWAC), acting as an informal group of agency representatives, focuses more narrowly on ground‐based gravitational wave networks. GWAC was an initiative of the US National Science Foundation (NSF) to open lines of communication among the various funding agencies world‐wide interested in the future of gravitational waves. It is open to any representative of a funding or national government agency, regardless of the current level of support provided to the field and with no financial commitment required. One potential hurdle for GWAC to overcome as a possible forum for the agencies funding a 3G network is that Japan’s MEXT (one of the major agencies funding current projects) and BMBF in Germany have not yet joined.\\
    \end{itemize}

How these two forums will evolve and whether they can develop common oversight and approval mechanisms for a 3G detector network is not yet clear. Similar organizations from high-energy physics, such as FALC, may serve as possible models.

Finally, we should note that there are signs of interest by important new agencies in becoming involved in the 3G network of gravitational wave detectors. The first generation of gravitational wave interferometers was funded by NSF (in the US), STFC (in the UK), MPG in Germany, CNRS \& INFN (Europe), and MEXT (Japan). As the second generation of detectors has come into operation, this has drawn in a number of additional participants, including NWO (Netherlands), ARC (Australia) and DST and DAE (India). Now, there have been indications of possible interest by CERN, US DOE and NASA, Helmholtz Association, Consejo Nacional de Ciencia y Tecnología (Mexico), the Canada Foundation for Innovation, and possibly others. The strength of the new interest and how these new players might be best integrated into the effort is still to be determined.

\clearpage
\chapterimage{Impact_pic.jpg} 

\chapter{Impact of Scientific/Technical Goals, \& Requirements}



{\Large\bf T}he governance model for the 3rd generation network must reflect its scientific and technical needs, and thus must be developed in parallel with the development of a strong science case and conceptual design for the network. In particular, it is essential to define the minimum number of detectors needed for the global 3G network with recommendations for location and orientation on the earth. This should be the outcome of vigorous scientific case studies and modeling.

Furthermore, in the case of many other multi‐site projects such as the Square Kilometer Array (SKA) and the Cerenkov Telescope Array (CTA), the science being carried out at each site tends to be different and self‐contained; such projects are often served by governance structures that give substantial independence to the different sites. We believe that this is unlikely to be the case for 3rd generation gravitational wave detector systems, although a case might be made for a dedicated stand‐alone detector system.

In preparation of this report, we conducted a Strength-Weaknesses-Opportunities-Threats (SWOT) analysis of 18 large infrastructures of nuclear, particle, astroparticle and astrophysics. It is also important to note that some of the authors of this report have or are still participating in the governance bodies of these infrastructures. The specific analyses are collected in an internal document, available on demand. The findings inform the analysis of challenges facing the 3G Gravitational wave program. For obvious reasons we do not provide the details of the analyses of each infrastructure, since these are inevitably subjective and the documentation is extensive quite long and goes beyond the mandate of the present committee. It is, however, useful in assisting further detailed studies.

Below, we list the four (4) challenges that will be faced along the way to 3G construction, not in a priority order but rather in a chronological order, describing challenges that will have to be faced earlier than others. Given their strong science content, a full evaluation is needed in discussion with the scientific community, and has to be addressed in detail in the coming years. These are a) the phasing in and out of the 2G and 3G infrastructures, b) the R\&D and policy towards the new instrumentation of 3G and 3G data treatment and access policies. 4) and the proper phasing of observation vs upgrade efforts during the operations phase.\\

\noindent{\textbf{\large{Challenge  1: The phasing of the transition from  2G to 3G}}}\\
In the transition from first generation detectors to the second (current) generation, the sensitivity gain was sufficiently dramatic to warrant ending operation of the earlier systems as scientifically inefficient, but because the second generation detectors were physically located at the sites of the first generation ones, continuity of the operation of the sites was easy. We envision a similar dramatic sensitivity and capability increase in the transition to the 3G network and we, therefore, assume that earlier systems will wind down their operations in   favor of the far more effective 3G systems. Nevertheless, an important point that has to be addressed along the road to 3G infrastructures is the rhythm of implementation of the already planned upgrades, since:

\begin{itemize}
\item they provide a clear asset for the field providing a unique opportunity to interleave in an orderly fashion, observation, technical upgrades and 3G design and construction,
\item they serve as a testing ground for 3G technology reducing the risks in the final implementation,
\item they will provide an increase of the number of detected sources by factors ranging from 10 to 30, including the unknown. They could therefore have an as yet unknown impact on the scientific targets that will guide the 3G design.
\end{itemize}

In any case, towards the end of the 20’s, or sometime in the next decade, the transitioning out of the current detectors, and their operations, in favor of the far more effective 3G systems will be likely. Current ideas for 3G detectors require new facilities and new sites. The leaders of the current projects and the broader gravitational wave community should carefully consider the optimal way to phase out the existing sites, and to smoothly transition both operational knowledge and personnel to the new facilities. The history of retiring large physics and astronomy facilities shows that this is a very complex and difficult task. Scientific, cost and political considerations will be significant in driving these decisions.

For the present we will assume that a minimum of 3 third generation (3G) detector sites are likely to be required, possibly with one in the USA, one in Europe and one in Asia or the Southern hemisphere.

Considerations of a sufficient baseline should be important here, mindful of geopolitical considerations. Additional nodes would be of substantial benefit depending upon separation, orientation, and site design. We also assume that the arm lengths will be substantially longer than the existing sites, thus requiring either new locations or such substantial expansion of the existing sites that the existing facilities do not strongly constrain the designs. It should be noted that, given this consideration, the civil and vacuum system costs are very likely to dominate the implementation costs for the 3G nodes.\\

\noindent{\textbf{\large{Challenge 2: Uniformity of infrastructure design vs diversity of technologies}}}\\
A major question to be answered is whether or not the detailed design of the detectors needs to be the same. Experience with LIGO shows the advantages of having a common design in terms of identifying and overcoming technical issues at multiple sites, but it has to be remembered that the pathway to achieving this common design was not simple. With many more countries involved in the design of third generation instruments, the situation will certainly be more complex, although the increasing maturity of the field may help to counter this.

The single most important driving factor towards a common design is the unity of the science case. If the science case is the same across sites, a common approach has obvious advantages. If there are strong and distinct diverse science cases displacing the broader common science goals, individual detectors may be focused on these. However, the global network, apart from specialized detectors or upgraded 2G detectors, must be strong enough to address the broad discovery science accessible in the 3G era. A robust global network almost certainly must be broadly capable.

One way to encourage a move towards a common design is to organize collaborative technology development in the major research groups around the world. This could be an essential part of the collaborative development that we recommend later in this report. In fact, the current regionally‐based 3G development efforts may best be viewed as the opening ideas in a broader, more comprehensive global development effort intended to yield the optimum 3G network. We describe a transitional approach to this later in this report.

Given the high cost of construction of and operating the third generation detectors, we believe that if the same design would be adopted for all the detector systems, with the same technical specifications and equipment requirements, design, construction and operating costs could be minimized. However, accommodating local site choices or the regional or local selection processes may justify diversity in the implementation. We do propose that the developmental program should be coordinated in a transitional organization to promote focus and prioritization and to facilitate an orderly process in making these choices.

It is essential that such an array is truly global in ownership, management and accessibility to data. For the sake of technical and purchasing efficiency it is also essential that, to as large an extent as possible, the underlying components of the facilities and the detectors are uniform, subject to constraints of availability or local conditions in different parts of the world. This approach will necessarily require the community to make significant compromises and down‐select decisions, as has been done on many other large international efforts.\\

\noindent{\textbf{\large{Challenge 3: Framework of data distribution, including low latency alarms, analysis and data access}}}\\
It has been gained through painful experience that, even in the 2G observatories, an early uniformity of the computing framework, including the distribution of alarms, is a prerequisite for the fulfillment of the global ambition of GW networks. There is currently a lot of work going on in the formation of the International Gravitational Wave Observatory Network (IGWN) on developing common cross- collaboration computing platforms. These efforts should be intensified towards closer coordination and uniformity of the computing frameworks. Here the experience of CERN, and the CERN experiments, can also be an asset. A second sensitive item in the data area is the open access issue. Agencies in the US and in Europe demand an open data policy and the availability of data as soon as possible, as a prerequisite for funding. This policy, fully justified from the citizen’s point of view, presents the danger that it is used as an opportunity for eventual funding partners, to avoid participation in the construction costs. The examples are many and well known (e.g. Vera Rubin Observatory). The proper balance has to be found and this is a challenge that has to be addressed at a global level. The practice of defining a proprietary period of around one year before full dissemination of data is a solution that has been adopted since, at least, the time of FERMI satellite, and it has been proven successful.\\

\noindent{\textbf{\large{Challenge 4: The phasing of observation periods vs periods of upgrade during the operations period}}}\\
Here again, this phasing is a common attribute of 2G GW detectors, as frequently the upgrades can yield better sensitivity resulting in dramatically increased event rates, and therefore have an advantage over a continuous operation as in standard electromagnetic sensing telescopes. The proper balancing is a scientific and technical problem, but has important governance impact, since it determines the corresponding roles of the on-site personnel and the collaboration at large. These roles go beyond the stereotypes of service-personnel and user-community and demand a flexible and resilient organization, with clear respective roles. It also has an impact on the quantity and quality of the personnel working on site, on the operational budget, and on the definition of full member agencies in the infrastructure vs partner agencies simply accessing the data of the infrastructure. This can become the distinguishing feature of a full member; that is participating at the definition of operation and upgrade policy, and therefore at the corresponding budgets of M\&O and upgrades, versus the partner agency, only participating in the exploitation of the data, paying only for the M\&O, including data curation.

\clearpage

\chapterimage{Role_of_Govnce_pic_v4.jpg} 
\chapter{The Role of Governance in Preparing a Successful Proposal}


{\Large\bf T}he envisioned 3rd generation gravitational network will be technically ambitious. It will serve as the means by which gravitational wave astronomy and associated multi‐messenger astronomy will become a mature exploratory science. It will revolutionize astrophysics. Its infrastructure and technical base will also be the platform for future evolution and progress in the field. The nodes in the 3rd generation network will not merely be one‐shot experiments. These nodes will be designed and implemented to a demanding standard and for a very demanding scientific mission.

The 3G network will also be very expensive. The current 10 kilometer underground Einstein Telescope (ET) concept, and the 40 kilometer surface-installed Cosmic Explorer (CE) concept, can easily \textit{each} be a few billion dollar/euros construction project. A network costing several billion dollars may only be feasible with global cooperation. For comparison, total costs for R\&D, design, construction and commissioning of these then fall into the cost ranges seen for the SKA implementation at the SKA South African and Australian sites. (The envisioned full implementation of SKA is estimated at several billion euros.) Another relevant example to help set the scale of investment is the CERN Large Hadron Collider including the two main detectors, ATLAS and CMS. These have required multiple billion dollar investments. But the CERN LHC complex is a singular installation. Other relevant examples are ITER, ALMA, ELT, TMT, ESS, ILC.

Can the 3G network be based upon individual nodes that, by themselves, may cost several billion dollars? Is it possible that a selected single node design as expensive as several billion dollars may cause only one new node to be supported in the next generation? If that were to be the outcome, it may prevent a true 3G network from being realized, leading to a single very powerful detector and remnants of the upgraded 2G detectors providing the network. Already, the experience of transitioning to the 2G network currently in place demonstrates the power of balanced, comparably sensitive, nodes in the network. The gravitational wave community must carefully consider the development of the science case and the definition of an affordable and balanced global network. A balanced network costing several billion dollars with several nodes delivered may be feasible with the appropriate global cooperation and coordinated global value engineering.

Furthermore, capital costs generate comparable scale lifecycle operating costs. The total investment in LIGO, from 1992 capital start through Advanced LIGO, including operations costs, constitute more than a billion dollar class investment to date. The gravitational wave community has ample experience in capital and operating expenditures in its multiple projects to date and these will guide planning for operational resources.

The examples of SKA and LHC additionally suggest the kind of community road mapping, R\&D, and development of funding agency consensus, as well as scientific, technical and project review, required to support decisions to proceed with the elements of a global third generation gravitational wave network. SKA antennas for various radio bands, and collider detectors for LHC, went through successive R\&D phases from early concepts to mature competing candidates and to precursor implementations in some cases. This progression took place on a base of existing mature large laboratories in these fields. The gravitational wave field is less developed than the radio astronomy and high‐energy physics fields that initiated SKA and LHC. It is reasonable to expect that the 3rd generation gravitational wave project will have to be developed with a similar progressive and thorough process.

The cost of a GW‐3G detector will be very substantially driven by the costs of the civil infrastructure. Furthermore, in the domain of civil infrastructure, including excavation, vacuum pipes and cryogenics, there are the same challenges and obvious synergies with proposals for the next generation colliders or megaton neutrino detectors. The adequacy of the computing infrastructure to address the increased flow of data is another challenge that has to be addressed, a challenge that is similar in nature with the challenge faced by SKA and even HL‐LHC. Given the current international experience in 3 regions of the world, one can derive scaling relations for construction and operation that obviously demand coordination and the implementation of innovative solutions. It is also essential that such an array is truly global in accessibility to data issues and therefore coordination needs to be extended to computing issues.

The governance structures for SKA and LHC each emerged from substantial collaboration between project leaders and funding agencies over extended periods of time. For LHC, the existing CERN intergovernmental organization, with its CERN Council and member funding agency consultations provided the needed confidence to adopt the recommendations of the scientific community. Future consideration of an international linear collider has similarly been considered by funding agency consultations initially through an ad hoc organization, Funding Agencies for the Linear Collider (FALC). For SKA, a similar funding agency consultative process turned community desires for the SKA into a progression of preparatory and design phases. These phases were carried out as international consortia, followed by a headquartered project operating as a member international company, and SKA has now evolved into an intergovernmental organization like CERN for the capital and operating phases of the SKA project. These interagency consultative processes have guided and overseen an exhaustive and progressive development of the early concept designs through R\&D into construction‐ready fully developed designs and project plans.

LIGO/Virgo, LHC and SKA are global efforts. The large resource streams required for these efforts are derived from the international partners. In these large communities, there is an increasing role for developing countries. Global scientific megaprojects have become pathways for economic and human capital development, commensurate with the large intellectual resource bases needed for these challenging efforts. This aspect of global science projects also drives the need for appropriate international consultation and coordination. As with radio astronomy requiring large radio‐quiet areas for array sites, the next phase of gravitational wave observations will require deployments in broad coverage around the globe, in seismically quiet sites, to assure source localization and to support sensitivity to polarization. Thus, scientific and resource imperatives drive global consultation and coordination.
Given these considerations, any proposals to sponsors for a 3rd generation gravitational wave detector network, and for each of the network nodes, will have to stand upon thorough scientific and technical development of the detector designs, as well as thorough development of the cost/schedule and project management infrastructure to support successful implementation. Careful evolution, confident designs and approaches and serious vetting will be required to support coordinated consensus for international commitments to proceed.

In this light, it should be assumed that the current 3rd generation interferometer concepts are to be viewed as the early stages in the development of the final adopted technical approaches which may be quite different from current concepts. As these technical developments progress, the project management, resource needs and governance should be defined in stages of analogous evolution, departing from the current collaboration of collaborations to a more appropriate and stronger management and governance model.

\clearpage
\chapterimage{Endpoints_pic.jpg} 
\chapter{Recommended End Point}


We have considered a wide range of governance models, tabulating their strengths and weaknesses, and evaluating their applicability to the specific nature of a 3rd Generation gravitational wave network. We believe that the following three structures are the best models for the ultimate governance for the construction and operations phase. Without specific recommendations at this time, these are:\\

1. International non‐profit member company such as a Delaware LLC, German GmbH, Dutch stiftung, etc. Companies of this type are fully recognized internationally. Governments can be members of the corporation and can make written company formation documents and contribution agreements that are legally binding and internationally recognized. These can have independent and appropriate procurement and employment systems and are fully responsible for all financial matters. Privileges and immunities enjoyed by the two international structures described below can be bestowed upon an international company by voluntary agreements with host countries as has been done in several cases in Chile and Spain.\\

2. An international research infrastructure consortium similar to the European Research Infrastructure Consortium (ERIC) model in which each partner deposits a letter of commitment to the agreed work of the consortium, signed by an appropriate government official. In the European model, the deposited commitment letters and research infrastructure agreement documents are deposited with the European Commission. Since the gravitational network considered here extends beyond a European perspective, we assume that a mechanism to implement an ERIC‐style structure in a broader international context can be implemented, by depositing the agreements in a broader international repository. We call this an International Research Infrastructure Consortium (IRIC). The commitments are voluntary by each government and can be changed by depositing a revised commitment letter. Carrying many of the features and strengths of an Intergovernmental Organization, this structure is less challenging to implement.\\

3.  An Intergovernmental Organization (IGO) based upon a treaty‐strength international Convention as exemplified by CERN or the recently formed SKA IGO. In this model, governments sign the fixed Convention in a very durable, powerful and inflexible commitment. We believe that forming a new intergovernmental organization for the 3G phase, and beyond, would be difficult to carry out though quite advantageous. However, an existing intergovernmental organization may choose to host the gravitational wave global effort. For this reason, we include this possibility here.\\

Note that all three of these models contain a strong central management. The final selection of which model to adopt must be made by the project leaders and the funding agencies together. The path to this ultimate governance structure described below is very similar no matter which of these three is chosen.

The current collaboration of collaborations, coordinated through individual agreements and through GWIC, should evolve into the chosen final governance model by the time construction commences. There are several examples of phased development of the governance structure paralleling the technical development, such as the aforementioned development of the SKA project. Given these examples, and based upon our view of current arrangements in the gravitational wave community, we propose the following evolution and suggest possible timelines for this process.\\

\begin{enumerate}
    \item{\textbf{Transitional period (2021-2023)}}\\
An initial transitional step, the current collaborations, possibly through GWIC but almost certainly requiring significant funding agency guidance and mandate, promptly forms a collaborative non‐legal “company”, governed by a collaborative board that guides actions by the legal sponsor/implementing agencies, to define a program of collaborative research and development, and design studies that progresses beyond the current ET and CE concepts. The ET and CE R\&D efforts should focus on developing technical feasibilities of selected technologies, and should evolve into exploring suitability for a broad range of sites.\\

The company should develop:
    \begin{enumerate}
        \item A detailed science case document, including multi‐messenger topics.
        \item An international collaboration on R\&D and civil infrastructure issues: mirrors, coating, quantum and cryogenic technologies, vacuum technologies, environmental monitoring and control, civil infrastructure technologies (e.g., excavation)
        \item A common framework to tackle the computing issues, including low latency and alert distribution.
        \item A common engaged roadmap from the current generation from 2G to 3G.
        \item Governance documents for the next phase together with involved funding agencies.\\
    \end{enumerate}

In addition it would coordinate:
    \begin{enumerate}
        \item Science and technical requirements for a third generation array of detectors
        \item Detector architectures including the level of design homogeneity that can be achieved in order to simplify the implementation, taking into account local conditions.
        \item Cost studies for site infrastructure and detector architectures and identification of value engineering priorities and cost cap targets.\\
    \end{enumerate}

Ideally, this program can be accomplished in 2 years from agreement to proceed recognizing that GWIC has already initiated some of the early stages of this process. After this 2 year period, one would expect:
    \begin{enumerate}
        \item A complete science case document for the 3G array
        \item Unique or coordinated solutions for the interferometer technologies and infrastructures, including computing
        \item A series of conceptual design reviews (CDRs) or technical design reviews (TDRs) on regional or global implementations, depending on the maturity of different regional initiatives, including site selection documentation and eventual choices
        \item Conceptual or detailed cost estimates and schedule definition
        \item Corporate documents as proposals for the next governance phase of coordination.\\
    \end{enumerate}

In fact, this phase represents “practicing” that type of governance by the participants, in a less formal framework.It should give way as rapidly as possible to the next type of organization, described below, that will continue the work above.
\newpage

\item{\textbf{A period of an interim international legal company (2023-2029)}}\\
A fully legal international company is formed as defined in the corporate documents drafted in the previous phase. This organization will be a member corporation with each of the sponsors/funding agencies providing representatives to serve on the governing board. This company will manage the developing project for the members who will pledge to contribute needed in‐kind and cash resources to this phase. The principle task in this phase is to do all work needed to bring the third generation project to construction readiness. During this phase, the baseline interferometers will reach final design, fabrication readiness, cost/schedule baselining and all needed elements of project definition. Interferometer and headquarters sites will be selected though all sites may not be equally ready for deployment and environmental impact processes may extend this step. Operations plans and costs will be defined at the conceptual level.\\

This phase should aim to be completed in a maximum of 6 years.\\

\item{\textbf{The final legal international company, or IRIC or IGO (2030-)}}, will be formed and first commitments to initiate construction will commence. This organization will manage the construction, transition to operations and will manage the operating observatory system.\\
\end{enumerate}

The key advantages of this program are the following:\\

\begin{enumerate}
    \item It immediately provides a coordinated framework for R\&D, infrastructure and computing solutions from the start, defining this task as a global issue. This is certainly a first with respect to previous history.
    \item It takes into account the different regional rhythms of transition, or funding opportunities, facilitating the progression of the GW field into a true global effort.
    \item It provides a solution for the diversity of members, since there are different organizational structures in every country. Even current consortia running 2G detectors, or other regional entities (e.g. ERIC) could eventually participate as separate entities. Other international organizations with common interests in R\&D and infrastructure could also participate as full or associated members.\\
\end{enumerate}

The progression described above brings the project to the construction stage in 6‐8 years from initiation. This process may well take longer than 8 years given the complexity of the technical, political and social goals. However, if construction on the first couple of interferometer sites commences in 6‐8 years and requires another 6‐ 8 years until first lock at these sites, based upon the initial LIGO (1994‐2001) and Virgo (1997‐2003) experience, the third generation will arrive in about 15‐16 years from a decision to initiate this phased progression. The current LIGO and Virgo interferometers, with envisioned running and upgrades should be in their final epoch at 15 years from now. Thus, the suggested timescales are appropriate goals for smooth growth of this exciting field.

This proposed evolution is in substantial resonance with the General Recommendations of the DAWN‐ IV meeting10 stating that: “a) GWIC should found an international Umbrella Organization by the Dawn V meeting in Spring 2019 to coordinate international research and development for 3G and detector upgrade plans; b) The ground‐based GW community should prepare to respond to calls for input to roadmaps. Specifically for the US Astro2020 Decadal Survey, through the submission of i) a coordinated set of science white papers, and ii) roadmaps for development of mid‐scale technologies and programs to enable 2.5 and 3G detectors. A proposal should also be included in the 2021 European Strategic Forum for Research Infrastructures (ESFRI) roadmap.”

Indeed, as we conclude this recommendation report, the ESFRI and Astro2020 surveys are well developed and their recommendations will emerge and may address this field. This governance report should be evaluated again upon receipt of the reports from these surveys. While these surveys are both regional in character, we reaffirm our recommendation that the 3G roadmap be approached in a global context given the revolutionary capability now evident and that the next step defines not just the third generation but the infrastructure path into the future for ground-based gravitational wave science.

\clearpage

\end{document}